\documentclass[twocolumn]{revtex4}

\def\be{\begin{equation}}
\def\ee{\end{equation}}
\def\la{\label}
\def\bea{\begin{eqnarray}}
\def\eea{\end{eqnarray}}
\def\non{\nonumber}
\def\ci{\cite}
\def\la{\label}
\def\bib{\bibitem}
\def\lesssim{{_ <\atop{^\sim}}}
\def\lm{\lambda}
\def\Lm{\Lambda}

\def\le{\left}
\def\ri{\right}

\def\Odmo{\Omega_{dm o}}
\def\Olo{\Omega_{\Lambda o}}

\def\Obmo{\Omega_{bdm o}}
\def\Obmc{\Omega_{bdm c}}
\def\Obo{\Omega_{bdm o}}

\def\Obm{\Omega_{bdm }}

\def\Obo{\Omega_{b o}}

\def\Oro{\Omega_{r o}}

\def\Omo{\Omega_{m o}}

\def\rbm{\rho_{bdm}}

\def\rbmo{\rho_{bdm o}}
\def\r{\rho}
\def\rc{\rho_c}
\def\ro{\rho_o}
\def\rcc{\rho_{core}}
\def\rrc{r_{core}}

\def\lfs{\lambda_{fs}}

\def\mfs{M_{fs}}

\def\s8{\sigma_8}

\def\fr{\frac}

\def\non{\nonumber}

\usepackage{graphicx}

\begin{document}

\title{BDM Dark Matter:  CDM with a core profile and a free streaming scale }

\author{A. de la Macorra}
\affiliation{Instituto de F\'{\i}sica, Universidad Nacional Autonoma
de Mexico, Apdo. Postal 20-364,
01000 M\'exico D.F., M\'exico\\
Part of the Collaboration Instituto Avanzado de Cosmologia}

\begin{abstract}

We present a new dark matter model BDM which  is an hybrid between
hot dark matter HDM and cold dark matter CDM, in which   the BDM
particles behave as HDM above the  energy scale $E_c$ and as CDM
below this scale. Evolution of structure  formation is similar to
that of CDM model but BDM predicts a nonvanishing free streaming
$\lfs$ scale and a inner galaxy core radius $\rrc$, both
quantities determined in terms of a single parameter $E_c$, which
corresponds to the phase transition energy scale of the subjacent
elementary particle model. For energies above $E_c$ or for a scale
factor $a$ smaller then $a_c$, with $a<a_c<a_{eq}$, the particles
are massless and $\r$ redshifts as radiation. However, once the
energy becomes $E\leq E_c$ or $a>a_c$ then the BDM  particles
acquire a large mass through a non perturbative mechanism, as
baryons do, and $\r$ redshifts as matter with the particles having
a vanishing velocity. Typical energies are
$E_c=O(10-100)\textrm{eV}$ giving a $\lfs\propto
E_c^{-4/3}\lesssim \textrm{Mpc}$ and $\mfs\propto E_c^{-4}\lesssim
10^9 M\odot$. A $\lfs\neq 0, \rrc\neq
0$ help to resolve some of the shortcomings of CDM such as
overabundance substructure in CDM halos and numerical fit to
rotation curves in dwarf spheroidal and LSB galaxies. Finally, our
BDM  model and the phase transition scale $E_c$ can be derived
from particle physics.

\end{abstract}

\maketitle

\section{Introduction}

The understanding of our universe has received a great deal
of attention in recent times. Cosmological data such as
large scale structure \ci{lss}, SN1a \ci{sn} and CMB \ci{cmb} are consistent
with the concordance $\Lm CDM$  model, with $\Odmo=0.22, \Olo=74,
\Obo=0.04$ with $h_o=0.71$. Even though the simple concordance model describes well
our universe the nature of dark energy DE and dark matter DM, which account for
up to $96\%$ of the energy content, is not well understood.
A large number of candidates
have been proposed for DM of which cold dark matter (CDM)
has been the most popular.  CDM model has been successful on
large scales in explaining structure formation in the
early universe as well as abundances of galaxy clusters \ci{lss}.
However, CDM predicts steeply cusped density profiles
and causing a large fraction of haloes to survive as substructure
inside larger haloes \ci{nfw,cdm-sub}. These characteristics of CDM haloes, however, seem to
disagree with a number of observations. The number
of sub-haloes around a typical Milky Way galaxy, as identified
by satellite galaxies, is an order of magnitude smaller
than predicted by CDM \ci{cdm-sub-probl} and the
observed rotation curves for dwarf spheriodal dSph and low surface brightness (LSB)
galaxies seem to indicate that their dark matter
haloes have constant density cores \ci{core-lsb,burkert} instead of steep cusps
as predicted by the NFW profile.
Low surface brightness
galaxies are diffuse, low luminosity
systems, with a total mass believed to be dominated
by their host dark matter halos \ci{dm-lsb}. Assuming that LSB
galaxies are in dynamical equilibrium, the stars
act as tracers of the gravitational potential, and can
therefore  be used
as a probe of the dark matter density profile \ci{bosh}.
Much better fits to dSph and LSB observations
are found when using a cored halo model \ci{core-mod}. Cored halos have
a mass-density that remains at an approximately constant
value towards the center.

It has been argued that the sub-structure and core problems might be solved
 in a NFW profile once additional baryonic physics are taken into account
such  as reionization and supernova feedback. This feedback
may help to suppress star formation and to decrease central densities in low-mass dark
matter haloes \ci{feedb}.
However, even tough these processes may help to solve the problem
with the over-abundance of satellite galaxies, the suggestion
that feedback processes can actually destroy steep
central cusps seems somewhat contrived in detailed simulations \ci{feedb-probl}.
Due to these discrepancies in CDM, numerous alternatives to
the CDM paradigm have been proposed. These include
broken scale-invariance \ci{bsi}, warm dark matter \ci{wdm},
scalar field dark matter \ci{sdm}, and various sorts
of self-interacting or annihilating dark matter \ci{sidm}.
However, these alternatives are unable to solve
both problems simultaneously.

Here we propose a new version of dark matter, well motivated from particle
physics, which predicts simultaneously a cut in the substructure formation
and cored center galaxies. The model simply consist of particles that
at high energy densities are massless relativistic particles with
a velocity of light, $v=c$, but at low densities they acquire a large mass, due
to nonperturbative quantum field effects, and become non relativistic with a  vanishing (small)
dispersion velocity. We will name this type of dark matter BDM, from bound states dark matter.
 The name is motivated by the particle physics model, discussed in section \ref{SPM}, but  we would
like to stress out that the cosmological properties of  BDM do not depend on this particle model
but on the different behavior of the BDM particles. The phase transition energy density is
defined  $\rc\equiv E_c^4$
and its value can be determined theoretical by the particle physics model or
phenomenological  by consistency with
the cosmological data.

There are two natural places where one encounters high energy densities for dark matter.
Firstly, at early times when
the universe is hot and dense, and secondly in the center of galaxies. From a cosmological
point  of view, our BDM has then two clearly distinct  behaviors: one as hot dark matter
HDM at high energies
$\rc<\rbm \sim a(t)^{-4}$ with $\Obm$ being constant and another as CDM for energies
smaller than $\rc>\rbm \sim a(t)^{-3}$.
The transition takes place at $a_c$ and
for $a<a_c$ the BDM  particles are relativistic but  for $a>a_c$ they become
non-relativistic. If BDM accounts for all DM than the redshift $a_c$, at which we have the
phase transition  $\r(a_c)\equiv \rc$, must be smaller than the
matter-radiation equality $a_{eq}$ and  $\rc \geq \r_{eq}$.
From  constraints on extra relativistic
energy density at the time of nucleosynthesis we will have an upper limit
for $a_c$ or equivalently a lower limit for $\rc$.

Since the particles of BDM travel at the
speed of light at energy densities above $\rc$ there
will be a cut at small scales in the power spectrum and BDM will erase inhomogeneities
and inhibit  structure formation for scales below the free streaming scale $\lfs$. This
property is similar to that of WDM and it predicts lower number of substructure in DM halos
as a CDM model, welcome by the data. The value of $\lfs $ will depend only on the phase transition
energy scale $E_c\equiv \rc^{1/4}$. Once the universe expands and $\rbm$ drops to values
smaller than $\rc$, the speed of the BDM particles vanishes and these particles will then be CDM
until present time.

Once DM dominates the universe, structure formation is effective, our
BDM particles are cold and we expect a standard CDM inhomogeneities growth but with
the a cutoff $\lfs$ in the power spectrum. The average energy densities in halos
is of the order $10^{5}\r_o$  and  as long as $\r_g<\rc$ we expect a standard CDM
galaxy profile, which may be given by the NFW profile.
However, once we approach  the center of the galaxy the energy
density increases and once it reaches the point $\r_g=\rc$  we encounter
a phase transition for the  BDM particles and they become massless again. We identify this energy
density of the galaxy with the core energy density $\rcc=\rc$ at a core galaxy radius  $r=\rrc$.
Inside $r<\rrc$ the
BDM particles are relativistic and
the DM energy density inside the radius  $\rrc$ remains constant avoiding a galactic cusp. Of
course we would expect a smooth  transition region between these two distinct behaviors but the effect
of considering the thickness of this transition region is a small and we will not consider it here.
As for the free streaming scale $\lfs$,  the size of $\rrc$ depends only on $E_c$.

Therefore, our BDM model predicts a free streaming scale $\lfs$, a cut off in the power spectrum
and a galaxy core radius $\rrc$  all derived in terms of a  single
parameter $E_c$. The value of $E_c$ is determined by the particle physics model and constraint by the
 dark matter properties derived from
cosmology.
Furthermore, since $E_c$  gives the phase transition scale of a particle
physics model further phenomenological consequences could arise. We present in section \ref{bdm}
the cosmological relevant properties of BDM model and in section \ref{SPM} the particle
physics model giving BDM. Finally, we present in section \ref{con} our conclusions.

\section{BDM Cosmology}\la{bdm}

We consider a standard FRW model universe with dark energy and dark matter. The dark matter
proposed here is motivated in section \ref{SPM} from particle physics and it
has a phase transition at an energy  scale $E_c$ or equivalently at an energy density  $\rc\equiv E^4_c$.
For energy densities larger than $\rc$ the particles have a peculiar velocity
$v=c=1$ and behaves as relativistic dark matter while for $\rbm<\rc$ the velocity vanishes
and $\rbm$ behaves as cold dark matter. We encounter naturally two regions in cosmology with a high
$\rbm$. The first one is at early times when the universe is hot and as long as  $\rbm(a)>\rc(a_c)$,
where $a_c$ is the transition redshfit, and the second
region is in the center of galaxies for a radius $r<\rrc$ with $\rbm(r)>\r(\rrc)=\rc$.

The cosmological evolution of the BDM energy density
and the peculiar velocity   in  terms of  the scale factor $a(t)$ is  then given by
\bea\la{rb}
\rbm(a)=\rc\le(\fr{a_c}{a}\ri)^{4}\geq \rc &,\;\;\;v=1,\;\;\;  a\leq  a_c\\
\rbm(a)=\rbmo\le(\fr{a_o}{a}\ri)^{3}\leq \rc &,\;\;\;v=0,\;\;\; a
\geq a_c
\eea
where the subscript $o$ is at present time and
\be\la{rbc}
\rc\equiv E_c^4 \equiv \rbm(a_c)
\ee
\begin{figure}[ht!]
\includegraphics[width=6cm]{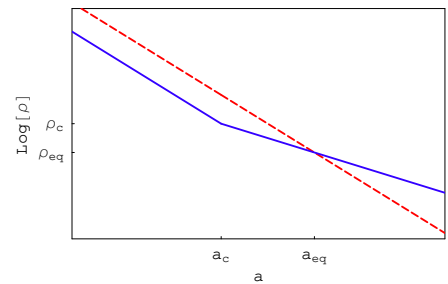}
\caption{\small{We show the behaviour of BDM and radiation (solid
(blue) and dotted (red), respectively) and the transition scale
$a_c<a_{eq}$ where BDM changes its evolution from $\r\propto
1/a^3$ to $\r\propto 1/a^4$. }} \label{fig2}
\end{figure}
\begin{figure}[ht!]
\includegraphics[width=6cm]{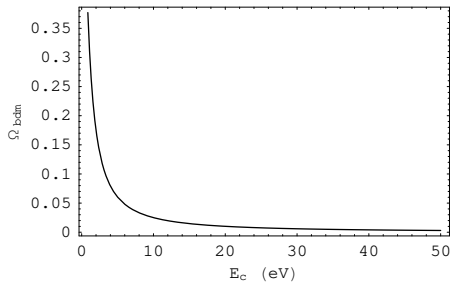}
\caption{\small{We show the dependence of $\Obm$ as a function of
$E_c$. $\Obm$ is smaller for larger $E_c$ and it is constant for $a<a_c$.}} \label{fig1}
\end{figure}
is the critical or
phase transition energy density. In terms  of a convenient phase
transition scale $E_c=O(10-100)\,eV$, derived from the particle
physics model  \ci{axel.mod} (c.f. sec.\ref{SPM}), we have
\be\la{rc}
\rc=\le(\fr{E_c}{60\,eV}\ri)^4 10^{17}\r_o =4\times 10^{10}
\le(\fr{E_c}{60eV}\ri)^4M_\odot/pc^3.
\ee
If we require $\rbm$ to
account for all dark matter the scale factor $a_c$ must be smaller
than $a_{eq}$, the matter-radiation equivalence scale factor. Of course it also possible that $\rbm$
accounts only partially for DM  in which case $a_c$ could be larger than
$a_{eq}$. We can
expressed $a_c$ in terms of the transition  energy density as
\be\la{ac}
a_c=\le(\fr{\rbmo}{\rc
}\ri)^{1/3}=\le(\fr{3H_o^2\Obmo}{E_c^4} \ri)^{1/3}
\ee
where we have
taken $a_o=1$, $\Oro$ is the relativistic energy density and
$\Omo=\Obmo+\Obo$
 the total matter at present time and $a_{eq}=\Oro/\Omo$. For $a_c \leq a_{eq}$,
 eq.({\ref{ac}) gives $E_c > 0.9\, eV$ with $E_{eq}=0.5\,eV$ and
$\Obm(a)$ remains constant for $a<a_c$ since the universe is dominated by radiation and
$\rbm$ behaves in this region also as a relativistic fluid. Therefore,
$\Obm(a\leq a_c)=\Obm( a_c)\equiv\Obmc$
is constant and in terms of the transition scale we have
\bea\la{ob}
\Obmc &=&\fr{\rc}{3H^2_c}=\fr{\Obmo}{\Omo}\le( \fr{q}{1+q}\ri)= 0.005\,\le(\fr{60 eV}{E_c} \ri)^{4/3}\non\\
q(E_c)&\equiv & \fr{a_c}{a_{eq}}=\fr{\Omo}{\Oro}\le(\fr{\rbmo}{\rc }\ri)^{1/3}=
 0.004\le(\fr{60\, eV}{E_c}\ri)^{4/3}
\eea
with
$3H^2_c=\r_{Tc}=\r_r(a_c)+\rbm(a_c)+\r_b(a_c)$,   $\Obmo/\Omo=0.22/0.26$ and  $q\sim 10^{-3}\ll 1$. The
amount of extra energy density is constraint by nucleosynthesis "NS".
The upper bound is $\Omega_{extra}(MeV)<0.1-0.2$ \ci{NS} and from eq.(\ref{ob})
we have then the constraint $E_c> (3.3-1.8)\,eV$ and
$q=a_c/a_{eq}< (0.13-0.3)$, respectively. We see that NS constraint allows the
transition redshift $a_c$ or $E_c$ to be quite close to the matter-radiation
equivalence values but with $E_c>E_{eq}$ and $ a_c<a_{eq}$. We show
in fig(\ref{fig1}) the evolution of $\r(a)$ with the pase transition
at $a_c$.  In fig.(\ref{fig2}) we show the dependence of $\Obm$ as a function
of $E_c$ for $a<a_c$.

It is well know that a HDM model has a cut in the power spectrum and does not allowed
structure to form below the free streaming scale $\lfs$  containing a mass  $\mfs=4\pi\ro (\lfs/2)^3 /3$.
A large $\mfs> 10^{12} M_\odot$, as for neutrinos, is ruled out by structure formation
but a smaller $\mfs\simeq 10^9 M_\odot$ would help to solve the problem of having too much
substructure in a CDM scenario.  The free streaming scale
$\lfs$ is given by
\bea\la{lfs1}
\lfs(t)&=&a(t)\int_0^{t}dt'\fr{v(t')}{a(t')}\non\\
&=&a(t)\int_0^{a_c}\fr{da}{a^2H}=\fr{a(t)}{H_c a_c}
\eea
where we have used $dt=da/aH$ and that the velocity of the BDM is
given by $v=1$ for $a\leq a_c$ and $v=0$ for $a>a_c$.
Using eq.(\ref{ac}) with $3H^2_c(a_c)=\r_{Tc}(a_c)=\rc/\Obmc$
and eq.(\ref{lfs1})  we get a free streaming scale
\bea\la{lfs2}
\lfs(E_c)&=&\fr{\sqrt{\Oro}}{H_o\Omo}\;\fr{q}{\sqrt{1+q}}=
\fr{\rbmo^{1/3}\;E_c^{-\,4/3}}{H_o\sqrt{\Oro}\sqrt{1+q}}\\
&\simeq& \le(\fr{60\,eV}{E_c}\ri)^{4/3}0.4\;\textrm{Mpc}
\eea
and a contained mass within a radius $\lfs/2$
\be\la{mfs}
\mfs(E_c) \equiv \fr{4\pi\r_o}{3} \le(\fr{\lfs}{2}\ri)^3=
\le(\fr{60\,eV}{E_c}\ri)^4 5.1\times 10^9 M_\odot.
\ee
\begin{figure}[ht!]
\includegraphics[width=6cm]{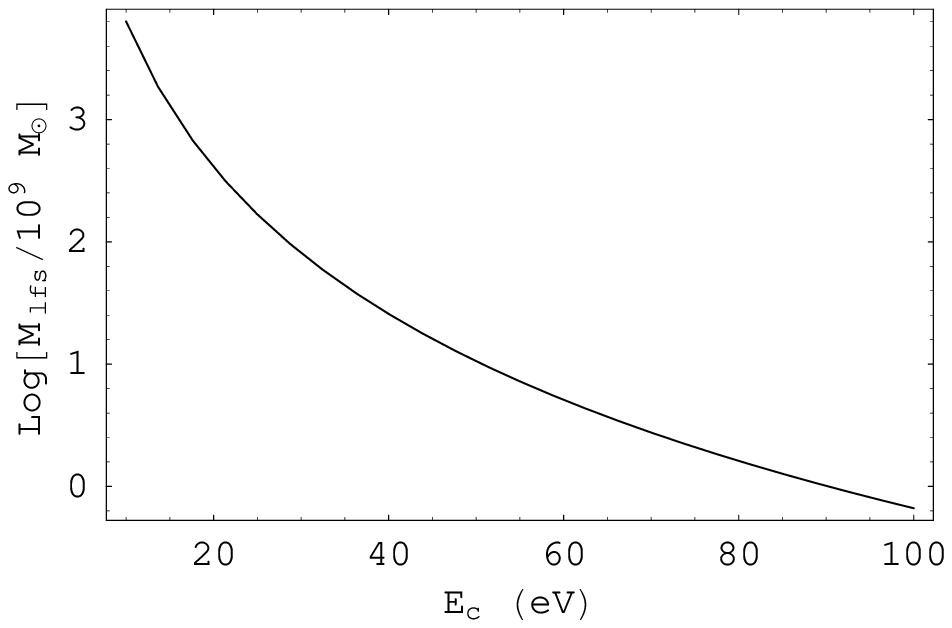}
\caption{\small{We show the dependence of $\mfs$ as a function of
$E_c$ and the larger $E_c$ the smaller $\mfs$. }} \label{fig3}
\end{figure}

\begin{figure}[ht!]
\includegraphics[width=6cm]{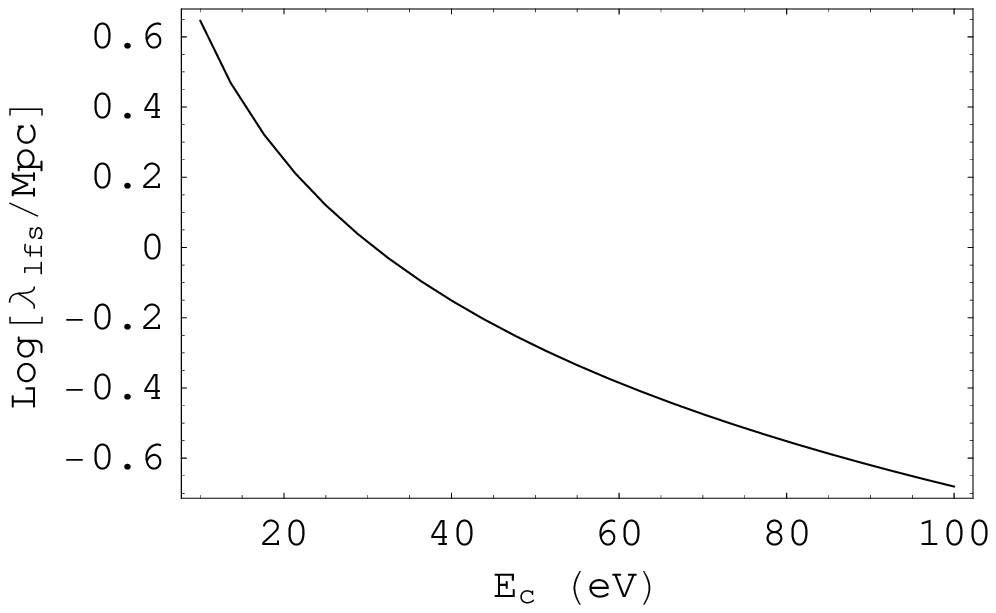}
\caption{\small{We show the dependence of $\lfs$ as a function of
$E_c$ and the larger $E_c$ the smaller $\lfs$. }} \label{fig4}
\end{figure}
We see that
$\lfs$ and $\mfs$ only depend on $E_c$ and are inversely
proportional to it. The larger the scale $E_c$ (or $\rc$) the
smaller $\lfs,\mfs$ will be. The value $E_c\sim 60 \,eV$ gives
precisely a $\mfs$ of the order of $10^9\,M_\odot$ as required to
inhibit extra substructure in halo but small enough to allow  for small
galaxies to be produced. For scales $\lm<\lfs$ there will be little power
on the spectrum. The constraint to have $\mfs<10^{10}M_\odot$ gives a lower
bound to $E_c$, with $E_c>50\,eV $ or  $\rc > (50\,eV)^4 $.
 In fig.(\ref{fig3}) and (\ref{fig4})  we show the dependence of $\mfs,\lfs$ as a function
of $E_c$.

In a spherical DM distribution the rotation velocity $vel$ can be
determine straightforward by $\nabla^2\Phi=4\pi\r$ with
$\Phi=-GM(r)/r$,  giving
\be\la{vel}
vel=\sqrt{\fr{G M(r)}{r}}
\ee
and $M=4\pi\int\r(r)\, r^2 dr $. The Jeans scale radius $r_J(E_c)=\lm_J/2$
is given by
\be\la{rj}
r_J(E_c)=\pi\sqrt{\fr{v^2_s}{4\pi
G\r}}=\sqrt{\fr{2}{3}}\fr{\pi}{E_c^2}= \le(\fr{60\,eV}{E_c}\ri)^2
11\,\textrm{pc}
\ee
where we have used  $\r=\rc=E^4_c$, the sound
speed for a relativistic fluid $v^2_s=1/3$ and $8\pi G=1$. The Jeans
scale implies that inhomogeneities at scales below $\lm_J$ inside a
galaxy are erased by the free streaming of the particles, i.e. we
have a constant inner galaxy core. The Jeans radius is the maximum
size for an inner core which coincides with the upper limit for a
core radius  $r \lesssim 300\,pc$ \ci{dm-lsb} giving a scale
$E_c\gtrsim  10 \,eV$. It is well known that in a cosmological scenario
the Jeans scale $r_J$ is larger than the Hubble radius $d_H=1/H$
for $a<a_{eq}$ and therefore structure formation does not form until
a matter domination universe. However, in a galactic scenario $r_J$ gives
the largest possible radius for a central core region.
\begin{figure}[ht!]
\includegraphics[width=6cm]{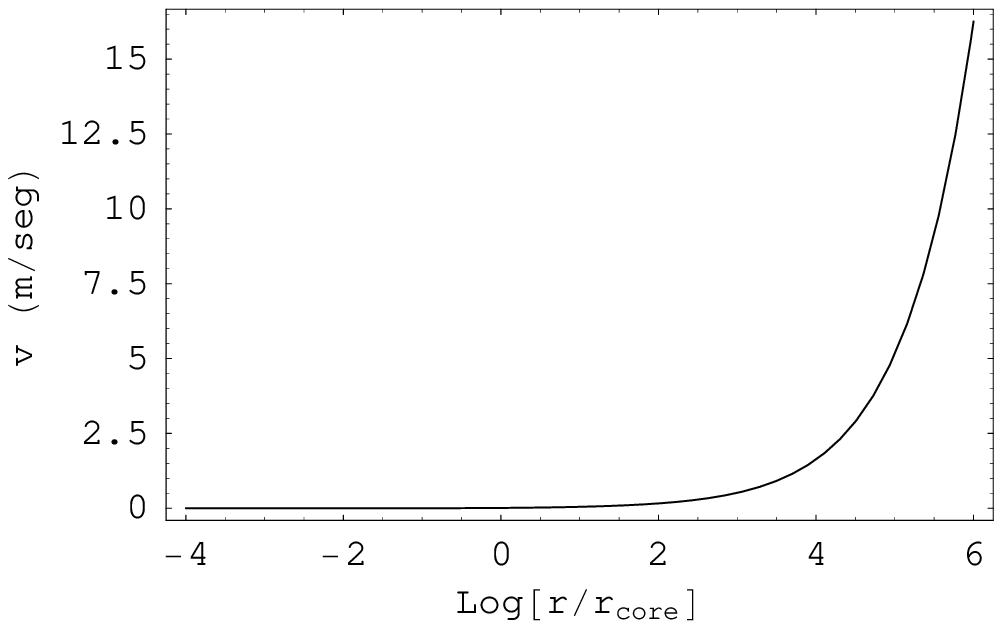}
\caption{\small{We show the inner circular velocity  of $v$ as a
function of $r/\rrc$ for a LSB type galaxy with
$\r_s=5\times 10^4\r_o,r_s=3h^{-1}kpc$. }} \label{fig5}
\end{figure}
\begin{figure}[ht!]
\includegraphics[width=6cm]{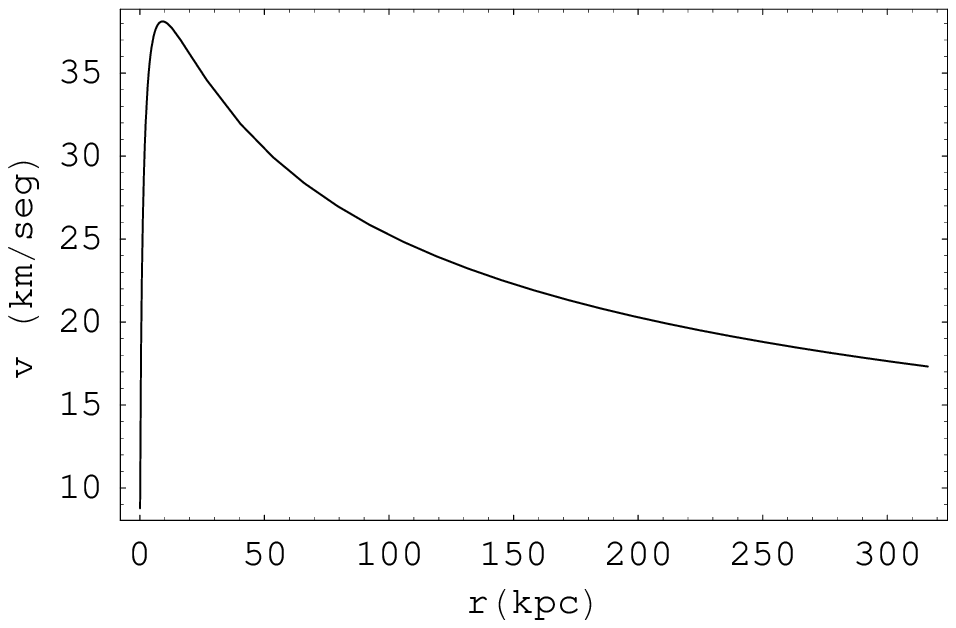}
\caption{\small{We show the circular velocity  of $v$
as a function of $r $ for a LSB type galaxy with
$\r_s=5\times 10^4\r_o,r_s=3h^{-1}kpc$. }} \label{fig6}
\end{figure}
Independent of the galaxy density profile we can estimate the
contained cored mass as a function of the core radius $\rrc$ using
eq.(\ref{rc})
\be\la{mc}
 M_{core}=\fr{4 \pi  \rc \rrc^3}{3}=40
\le(\fr{E_c}{60eV}\ri)^4 \le(\fr{\rrc}{0.001pc}\ri)^3M_\odot.
\ee
We see from eq.(\ref{mc}) that a contained mass
$M_{core}=O(1)M_\odot$ requires a small $\rrc=O(10^{-3})pc$ for a
$E_c\simeq O(10-100)eV$. Since $\rrc$ is small we expect a small
deviation from a CDM profile. There are different galaxy profiles
in the literature. The standard CDM profile is given by the NFW
$\r_{NFW}=\r_s/[(r/r_s)(1+r/r_s)^2]$ \ci{nfw} while a core galaxy
may be fit with $\r (r)=\r_b/[(1+r/r_b)(1+(r/r_b)^2)]$
\ci{burkert} with $\r_b,r_b$ the core density and radius,
respectively. The NFW profile has a cuspy inner region with
$\r_{NFW}$ diverging in the center of the galaxy with an inner
slope $\alpha=-1$ ($\r_g\propto r^{\alpha}$)  and $\r_s,r_s$ are
galaxy dependent parameters. For typical LSB galaxies one has
$\r_s\simeq 10^4\ro,r_s\simeq 3h^{-1}kpc$ \ci{nfw}. As mentioned
in the introduction, a core inner region with constant $\r$ and a
slope $\alpha<-1/2$ seems to be preferred  by dSph and LSB
galaxies \ci{dm-lsb}. The core  region is reached at $r\leq \rrc$
when the galaxy energy density reaches the value $\rc$,
\be
 \r_g(r)\simeq \rcc(\rrc)\equiv \rc.
\ee
The size of the core radius $\rrc$ depends on the
choice of the galaxy profile.
Since our BDM behaves as CDM for $\r<\rc$  as long
as the density of the galaxy is $\r_g< \rc$ we expect to
have a NFW type profile. Therefore,
a possible BDM profile is given by a cored CDM  profile as
\be\la{pbdm}
\r_g (r)=\fr{\r_s}{(\rrc/r_s+r/r_s)(1+r/r_s)^2}
\ee
with $\rrc\ll r_s$. This profile coincides with $\r_{NFW}$ at large
radius but has  a core inner region at  $r=\rrc$ with
$\r_g(\rrc)=\rc=\r_s r_s/2\rrc$   giving a core radius
\be\la{rco}
\rrc(E_c)\equiv \fr{\r_s r_s}{2\rc}=\fr{\r_s r_s}{2E_c^4}.
\ee
The slope is
\be\la{sl}
\alpha\equiv \fr{d\, Log[\r]}{d\, Log[ r]}=-\fr{r/\rrc(1+3r/r_s+2\rrc/r_s)}{(1+r/\rrc)(1+r/r_s)}
\ee
and
takes the values $\alpha=(0,-1/2,-1,-2,-3)$ for $r=(0,\rrc,\rrc\ll
r\ll r_s,r_s,r_s\ll r)$ respectively.
For values of $r\ll r_s$, eqs.(\ref{vel}) and (\ref{pbdm}) gives in terms of $r'\equiv r/\rrc$
a central mass, circular velocity and a slope
\bea
M(r')&\simeq &8\pi \rc \rrc^3\le(\fr{r'^2}{2}-r'+Log[1+r']\ri)\\
v(r')&\simeq &\rrc\sqrt{\rc} \le(\fr{r'}{2}-1+\fr{Log[1+r']}{r'}\ri)^{1/2}\\
\alpha(r')&\simeq &-\fr{r'}{1+r'}.
\eea
Notice that from
eqs.(\ref{lfs2}) and (\ref{rco}) we can express $\lfs $ in terms of
$\rrc$
\be
\lfs = b\;  \rrc^{1/3}
\ee
with $b=(2\sqrt{\rbmo}/\r_s r_s)^{1/3}/( H_o\sqrt{\Oro(1+q)})$
a proportionality constant, showing explicitly the interconnection of $\lfs$ and $\rrc$.
In fig.(\ref{fig5})  we show the central circular
velocity as a function of $r/\rrc$ while in fig. (\ref{fig6}) we plot the complete
circular velocity as a  function of $r(kpc)$. In fig.(\ref{fig7})  we show the central slope profile
as a function of $r/\rrc$ while in fig. (\ref{fig8}) we have the complete $\alpha$ as a function
of $r(kpc)$.

\begin{figure}[ht!]
\includegraphics[width=6cm]{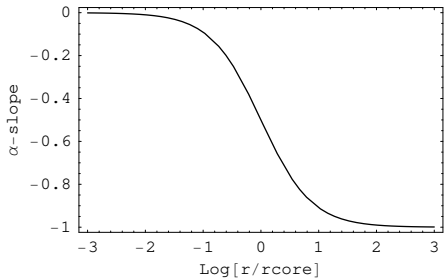}
\caption{\small{We show the inner slope of the galaxy profile in
eq.(\ref{pbdm}) as a function of $r/\rrc$ . }} \label{fig7}
\end{figure}

\begin{figure}[ht!]
\includegraphics[width=6cm]{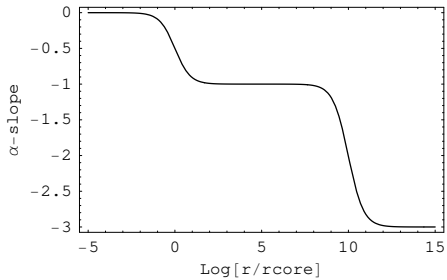}
\caption{\small{We show the complete slope of the galaxy profile
in eq.(\ref{pbdm}). }} \label{fig8}
\end{figure}

\section{Particle Model}\la{SPM}

In particle physics there are two different dynamically ways
to generate a particle mass, namely the Higgs mechanism and the
a non-perturbative gauge mechanism. In the SM the elementary  particles
(quarks, electrons, neutrinos) get their mass by the interaction with
the Higgs field. The dynamically evolution  of the Higgs field implies
that at high energies all SM masses vanish but once the Higgs settles
into the minimum of its potential, at the electroweak scale $E_{ew}=O(100GeV)$,
it acquires a non vanishing vacuum
value giving a mass to the SM particles. Therefore, the mass  of the
fundamental particles vanishes at high energies and are non zero
below the phase transition scale $E_{ew}$. On the other hand,
the non-perturbative gauge mechanism is based on the strength of
gauge interaction.  The strength of the coupling
constant "g" depends on the  energy as
\be\la{g}
g^{-2}(E)=g_i^{-2} + 8\pi^2 \,b\; \textrm{Log}\le[\fr{E}{E_i}\ri]
\ee
with $g_i=g(E_i)$ and $b$ counts the number of
elementary particles "Q" charged under the group. If
$b>0$, as for the strong force QCD with $SU(N_c=3), N_f=6$
($N_c,N_f$ are the number of colors and flavors,
respectively)
one has $b=(11N_c-2N_f)/3=7$, the gauge
coupling constant   increases with decreasing energy $E$
and we have a non-abelian asymptotic free gauge group.
The condensation or phase transition scale is defined as
the energy when the coupling constant becomes strong, $g(E)\gg 1$,
and from eq.(\ref{g}) we have
\be\la{e}
E_c= E_i\,e^{-8\pi^2/bg^2_i}.
\ee
The fact that $E_c$ is exponentially suppressed compared to $E_i$
allows as to understand why $E_c$ can be much smaller then
the initial $E_i$ which may be the Planck,
Inflation or Unification scale.
The scale $E_c$ sets a phase transition scale where above
$E_c$ the elementary particles "Q" (e.g. quarks in QCD) are
(nearly) massless and below $E_c$ the strong force binds
these elementary fields together forming neutral
bounds states such as mesons and baryons in QCD. The order of magnitude of
the mass of these particles is
\be\la{mb}
m_{BS}=d\,E_c
\ee
with $d=O(1)$ a proportionality constant and a Compton wave length
\be\la{lc}
\lm_{com}=\fr{1}{m_{BS}}=\le(\fr{60\,eV}{m_{BS}}\ri) 2\times10^{-5} cm
\ee
In QCD one has $E_c\simeq 200\,MeV$ with the pion mass $m_\pi\simeq140 MeV$
while the baryons mass (protons and neutrons) $m_b\simeq 940 MeV$,
i.e. the proportionality constant is in the range $0.7<d<5$, and
with bound mass much larger than the mass of the quarks ($m_u\simeq (1-3) MeV,
m_d\simeq (3.5-6) MeV$). Clearly the mass of the bound states is
not the sum of its elementary particles but is due to the non-perturbative
effects of the strong force and is well parameterized by $E_c$.
The dynamical formation of bound states is not completely understood since it involves non
perturbative physics. However, it has been shown in RHIC \ci{rhic} that
at high density, above the transition scale $E_c$, the QCD quarks
do indeed behave as free particles, while at low energies there are no
free elementary quarks and all quarks form gauge neutral bound states.
Since the interaction strength  increases at lower energies, the formation
of bound states is expected to be larger at the smallest possible
particle bound state energy $E_{BS}$ (i.e. $E_{BS}=m_{BS}$) with momentum $p^2=E_{BS}^2-m_{BS}^2
\simeq 0$. The energy  distribution of bound states formation is still under
investigation \ci{bound} and   for simplicity we take here $p=0$ which
gives a vanishing particle velocity for the bound states.

It is precisely the non-perturbative gauge mechanism that we have in mind for our
bound states dark matter BDM. Of course, in our case the gauge group
and elementary fields {\it are not} part of the SM. This "dark" gauge group is
assumed to interact with the SM only through gravity and is widely predicted
by extensions of the SM, such as brane or string theories. Furthermore,
this dark gauge group may also account for dark energy as dark mesons
\ci{axel.mod}. The model used in \ci{axel.mod} has a transition scale $E_c=O(10-100)\,eV$
which is the reference energy used here. We see from eq.(\ref{lc}) that for
$m=d\,E_c=60\,eV$ we have a small Compton wave length $\lm_{com}\simeq 10^{-5}cm$
and we therefore do not expect it to play a relevant role in late cosmology
or galactic scales.

Even tough we have motivated our BDM in terms of a well motivated
particle physics model we stress the fact that  the cosmological
implications of BDM do not depend on its origin of. The  BDM
is defined by a DM that at $\rbm>\rc$   behaves as relativistic HDM
with a particle velocity $v=1$ while for $\rbm<\rc$ it is
CDM with $v\simeq 0$. Naturally one encounters high $\r$ at early cosmological times and in the
inner regions of galaxies.

\section{Summary and Conclusions}\la{con}

We have presented a new type of dark matter, BDM, which can be derived
from particle physics. These new particles depend on a dark
gauge group, similar as the SM elementary particles, but with only gravitational
interaction with the SM. The BDM particles acquire a non perturbative mass
below the phase transition scale $E_c$. The order of magnitude
for this phase transition is $E_c=O(10-100)\, eV$ \ci{axel.mod}. Above $E_c$ we have relativistic
elementary particles while below this scale we have nonrelativistic bound
states, formed as neutral bound states from the elementary fields.
The cosmological effect of the phase transition
manifest itself in two different regions. Firstly,  in the cosmological evolution of the
energy density $\rbm$ and secondly in the inner regions of galaxies. From a cosmological evolution
we have a $\rbm$  redshifting as radiation above $\rc=E_c^4$ and as matter below
this scale. Since the BDM particles are relativistic at high energies we have
a nonvanishing  free streaming scale $\lfs$ with a contained mass $\mfs$. At the same
time our BDM predicts a core galaxy profile with radius $\rrc$ once the energy density of the galaxies $\r_g$
reaches the transition scale $\rc$. The quantities $\lfs,\mfs$ and $\rrc$ are given by eqs.(\ref{lfs2}), (\ref{mfs}) and
(\ref{rco}) and have the same origin, namely the phase transition scale, and are determined
in terms of a single parameter $E_c$. Therefore, we have proposed a DM model which connects
the solution to two of the main shortcomings of CDM, namely the overabundance of substructure in CDM halos
and the rotation curves for DM dominated galaxies.

\noindent{\bf Acknowledgment}--  We thank for
partial support Conacyt Project 80519, IAC-Conacyt Project.

\end{document}